\documentclass[aps,prb,twocolumn,superscriptaddress,showpacs]{revtex4}
\usepackage{amsmath,bm}
\usepackage{graphicx,epsfig}

\renewcommand{\v}[1]{{\bf #1}}
\newcommand{\be}{\begin{equation}}
\newcommand{\ee}{\end{equation}}
\newcommand{\nn}{\nonumber \\}
\newcommand{\ij}{\langle ij \rangle}

\newcommand{\ba}{\begin{eqnarray}}
\newcommand{\ea}{\end{eqnarray}}

\newcommand{\calAbar}{{\cal A}^\dag}
\newcommand{\calA}{{\cal A}}

\newcommand{\bw}{\begin{widetext}}
\newcommand{\ew}{\end{widetext}}
\newcommand{\la}{\langle}
\newcommand{\ra}{\rangle}
\newcommand{\ph}{\phantom}
\newcommand{\mk}{\bar{k}}
\newcommand{\bpm}{\begin{pmatrix}}
\newcommand{\epm}{\end{pmatrix}}

\newcommand{\kbar}{\overline{k}}

\def\mathlla[internal#1#2{%
 \\llap{$\mathsurround=0pt#1{#2}$}%
 }
\def\clap#1{\hbox to 0pt{\hss#1\hss}}
\def\mathclap{\mathpalette\mathclapinternal}
\def\mathclapinternal#1#2{%
 \clap{$\mathsurround=0pt#1{#2}$}%
}

\begin{document}

\title{Coupling of phonons and spin waves in triangular antiferromagnet}
\author{Jung Hoon Kim}
\affiliation{Department of Physics, BK21 Physics Research Division,
Sungkyunkwan University, Suwon 440-746, Korea}
%
%
\author{Jung Hoon Han}
\email[Electronic address:$~$]{hanjh@skku.edu}
\affiliation{Department of Physics, BK21 Physics Research Division,
Sungkyunkwan University, Suwon 440-746, Korea} \affiliation{CSCMR,
Seoul National University, Seoul 151-747, Korea}

\begin{abstract}
We investigate the influence of the spin-phonon coupling in the
triangular antiferromagnet where the coupling is of the
exchange-striction type. The magnon dispersion is shown to be
modified significantly at wave vector $(2\pi,0)$ and its
symmetry-related points, exhibiting a roton-like minimum and an
eventual instability in the dispersion. Various correlation
functions such as equal-time phonon correlation, spin-spin
correlation, and local magnetization are calculated in the presence
of the coupling.
\end{abstract}

 \maketitle

\section{Introduction}
A number of recent experimental breakthroughs has revived interest
in the phenomena of coupling between magnetic and electric (dipolar)
degrees of freedom in a class of materials known as
``multiferroics"\cite{review}. Some noteworthy observations include
the development of dipole moments accompanying the helical spin
ordering\cite{lawes,kenzelmann}, displacement of magnetic ions at
the onset of magnetic order in the triangular lattice
YMnO$_3$\cite{park}, and adiabatic control of dipole moments through
applied magnetic fields\cite{kimura,hur}, all of which unambiguously
point to the strong coupling of electric and magnetic degrees of
freedom. A number of theories has been advanced to establish a
microscopic understanding of these
couplings\cite{KNB,mostovoy,dagotto,harris,JONH,spaldin,jia-han}.

The known mechanisms of the spin-polarization coupling fall into two
categories. One is of the inverse Dyzaloshinskii-Moriya (DM) type,
whereby the local dipole moment, denoted by $u_{ij}$, couples to the
spins $S_{i}$ by $\sim u_{ij}\cdot \hat{e}_{ij}\times (S_{i}\times
S_{j})$. Here the unit vector $\hat{e}_{ij}$ connects the centers of
the magnetic ions at $i$ and $j$. The microscopic origin of such
coupling was investigated in, for instance,
Refs.~\onlinecite{KNB,JONH}. The behavior of a large class of
multiferroic materials can be understood on the basis of this
coupling\cite{review}. The other type arises from exchange-striction,
wherein the movement of the magnetic ions is assumed to directly
influence the exchange integral and lead to the coupling $\sim
\hat{e}_{ij}\cdot (u_{i}-u_{j})S_{i}\cdot S_{j}$. The spin-lattice
coupling in RMn$_{2}$O$_{5}$ (R=rare earth) is believed to arise from
this mechanism\cite{125}.

More recently, the dynamical aspect of the magnetoelectric coupling
has been investigated both experimentally\cite{pimenov,drew} and
theoretically\cite{KBN}. The dynamical coupling is important because
it can arise without the ordering of one or both of the degrees of
freedom, and can substantially influence the ac dielectric
response\cite{pimenov,drew,KBN}, or even result in an exotic new
phase with vector chirality\cite{onoda-nagaosa}. The dynamics of the
small fluctuations in the ordered phase of both the polarization and
the spin were examined in Ref.~\onlinecite{KNB} for the
one-dimensional frustrated chain. A corresponding analysis of the
coupled fluctuations has not yet been tried in the case of the
exchange-striction mechanism, or for other lattice geometries.

Triangular geometry offers a potentially fertile ground for the
interplay of spin and dipolar degrees of freedom because of the
tendency of spins to form a spiral ($120{^\circ}$) structure even
without further frustrating interactions. The ground state is
characterized by non-zero $\la S_{i}\cdot S_{j}\ra$ as well as $\la
S_{i}\times S_{j}\ra$, which may result in spin-dipole interactions
of both DM and exchange-striction types. Furthermore, quite recently,
it was shown that a spin $S=5/2$ triangular antiferromagnet in
RbFe(MoO$_4$)$_2$\cite{RFMO} develops spontaneous polarization along
the $c$ axis as the spins order in the $ab$ plane. This and another
triangular lattice compound, YMnO$_3$, offer promising examples where
the interplay of spin and dipolar degrees of freedom can be revealed
in detail. In particular, the dynamical aspect of the spin-dipole
coupling in the triangular lattice remains largely unexplored and a
theoretical consideration of their interplay would be timely.

In this paper, we examine the coupled dynamics of Heisenberg spins
and the local dipolar variable (hereafter referred to simply as
phonons) on the triangular lattice, assuming the exchange-striction
interaction. In Sec.~\ref{sec:spin-phonon}, the model Hamiltonian is
introduced and solved within Holstein-Primakoff theory. A number of
physical quantities, such as the local magnetic moment, phonon
correlation function, and the dynamic spin-spin correlation, are
derived in Sec.~\ref{sec:physical quantities}. Conclusions and the
relevance of our work to existing experiments can be found in Sec.
\ref{sec:discussion}.

\section{Spin-Phonon Model}\label{sec:spin-phonon}

The spin-phonon coupled Hamiltonian in the exchange-striction
picture reads\cite{jia-et-al}

\setlength\arraycolsep{2pt}
\begin{equation}
H \! = \!\sum_{\langle ij \rangle}[J_0 \!-\! J_1 \hat{e}_{ji} \cdot
(u_j \!-\! u_i ) ] S_i \cdot S_j + \sum_i \left( {p_i^2 \over 2m}
\!+\! {\frac{K}{2}} u_i^2 \right), \label{our-model}
\end{equation}
where the antiferromagnetic exchange integral $J_{ij}$ connecting
the two nearest-neighbor Heisenberg spins is expanded to first order
in the displacement $u_i$ of each atomic site $i$. $\hat{e}_{ji}$ is
the unit vector extending from $i$ to $j$ atomic sites. The terms
proportional to $J_1$ define the spin-phonon coupling. The
Heisenberg spin of magnitude $S$ is represented by $S_i$, and the
two-dimensional displacement vectors and their canonical conjugate
operators by $u_i$ and $p_i$. We separate the Hamiltonian into two
parts, $H = H_{0} + H_{1}$, where $H_{1}$ is the spin-phonon
interaction term and $H_{0}$ consists of the Heisenberg and phonon
Hamiltonians

\ba H_{0} &=& J_{0}\sum_{\la ij\ra}S_{i}\cdot S_{j}
+\sum_{i}\Big(\frac{p_{i}^{2}}{2m} + \frac{K}{2}u_{i}^{2}\Big), \nn
H_{1} &=& J_1 \sum_{\ij} \hat{e}_{ji} \cdot (u_i - u_j ) S_i \cdot
S_j . \label{eq:H}\ea
The classical ground state of the above Hamiltonian was worked out
in Ref.~\onlinecite{jia-et-al}. It was shown that, despite the
spin-phonon coupling term, the classical spin configuration is that
of the pure Heisenberg model on the triangular lattice with the
neighboring spins at a $120^{\circ}$ angle with each other.

Although the spin-phonon interaction does not produce any observable
effect in the ground state spin configuration, the excitation
spectra of the lattice (phonons) and the spins (magnons) will be
mixed due to $H_1$.

The small fluctuations near the ground state can be analyzed within
the standard Holstein-Primakoff (HP) approach after first rotating
the spin operators according to their classical spin orientations
defined by $\langle S_i \rangle = (0, \sin\phi_i, \cos\phi_i)$,
where $\phi_{i}$ is the angle the spin makes with the $z$-axis at
site $i$. All the spins are assumed to lie in the $yz$-plane, which
also coincides with the plane of the lattice itself. After
performing the Bogoliubov rotation defined by the angle
$\tanh\phi_{k} = 3\gamma_{k}/(2+\gamma_{k})$ to obtain the spin wave
spectrum, the Hamiltonian $H_0$ reads

\be H_{0} = \sum_{k}\varepsilon_{k}\alpha_{k}^{\dag} \alpha_{k}  +
\omega_0 \sum_{k} \Big( b_{ky}^{\dag}b_{ky} + b_{kz}^{\dag}b_{kz}
\Big)\ee
with the spin wave dispersion

\ba \varepsilon_{k} =
\frac{3J_{0}S}{2}\sqrt{(1-\gamma_{k})(1+2\gamma_{k})}. \ea
Here $\gamma_{k} = (1/3)\sum_{ \alpha =1}^3 \cos [k\cdot
\hat{e}_\alpha ] $ with $\hat{e}_1 = (1,0)$, $\hat{e}_2 =
(-1/2,\sqrt{3}/2)$, and $\hat{e}_3= (-1/2,-\sqrt{3}/2)$. The lattice
constant is taken to unity. Phonon operators in the $y$ and $z$
directions are also introduced above as $b_{ky}$ and $b_{kz}$ as
well as the phonon energy $\omega_0$.

The spin-phonon Hamiltonian $H_1$ can be expanded to second order in
the magnon and phonon operators. The full spin-phonon Hamiltonian to
quadratic order is given in the simple form ($\kbar\equiv
-k$)\cite{comment}

\ba  H &=& \sum_{k}\Bigg[ \varepsilon_{k}\alpha_{k}^{\dag}\alpha_{k}
+ \omega_0 \beta_{k}^{\dag}\beta_{k} + i\lambda_{k}
\Big(\alpha_{k}^{\dag} - \alpha_{\mk}\Big)\Big(\beta_{k} +
\beta_{\mk}^{\dag}\Big) \Bigg]\nn
\label{fullbogham}\ea
where

\ba  \lambda_{k} &=&
-\frac{3}{4}J_{1}S\sqrt{\frac{S}{m\omega_0}}e^{{\phi_k  \over 2}}
\sqrt{\chi_{ky}^{2} + \chi_{kz}^{2}}, \nn
\chi_{ky} &=& \cos(k\cdot \hat{e}_{3}) - \cos(k\cdot \hat{e}_{2}),
\nn
\sqrt{3}\chi_{kz} &=& 2\cos(k\cdot \hat{e}_{1}) - \cos(k\cdot
\hat{e}_{2}) - \cos(k\cdot \hat{e}_{3}) . \ea
Note that only the following linear combination of the phonons
participate in the interaction with the magnons:

\be \beta_{k} = \frac{\chi_{ky}b_{ky} +
\chi_{kz}b_{kz}}{\sqrt{\chi_{ky}^{2} + \chi_{kz}^{2}}}.\ee
The rotation of the Hamiltonian to the diagonalized basis is
effected by a series of canonical transformations given by $\psi_k =
W_k X_k Y_k \Psi_k $ where $\Psi_k$, the diagonal operators, are
given by $({\cal A}_{1k} , {\cal A}^\dag_{2\kbar} , {\cal A}_{2k} ,
{\cal A}^\dag_{1\kbar} )^T$. The respective matrices are defined as
follows:

\bw \ba W_k = \bpm \ph{-}\cos\frac{\theta_{k}}{2} & 0 &
-i\sin\frac{\theta_{k}}{2} & 0 \\
0& i \sin\frac{\theta_{k}}{2} & 0 & \ph{-}\cos\frac{\theta_{k}}{2}
\\
-i \sin\frac{\theta_{k}}{2} & 0 & \cos\frac{\theta_{k}}{2} & 0 \\
0 & \cos\frac{\theta_{k}}{2} & 0 & i\sin\frac{\theta_{k}}{2}
\epm,&\quad& \left\{ \begin{array}{c} \sin\theta_k =
\frac{2\lambda_k}{\sqrt{ (\varepsilon_{k}-\omega)^{2} +
4\lambda_{k}^{2}}} \\ {} \\ \cos\theta_k = \frac{\varepsilon_{k} -
\omega}{\sqrt{ (\varepsilon_{k}-\omega)^{2} + 4\lambda_{k}^{2}}}
\end{array} \right.\nn
X_{k} =
\bpm \cosh{\nu_{k} \over 2} & -i\sinh{\nu_{k}\over2} & 0 & 0 \\
i\sinh{\nu_{k}\over2} & \cosh{\nu_{k}\over2} & 0 & 0 \\
0 & 0 & \cosh{\nu_{k}\over2} & -i\sinh{\nu_{k}\over2} \\
0 & 0 & i\sinh{\nu_{k}\over2} & \cosh{\mu_{1k} \over 2} \epm,&\quad&
\left\{ \begin{array}{c} \sinh\nu_k =
\frac{\Delta_k}{\sqrt{\Delta_{k}^{2} - \Lambda_{k}^{2}}}
\\ {} \\ \cosh\nu_k =
\frac{\Lambda_k}{\sqrt{\Delta_{k}^{2} - \Lambda_{k}^{2}}}
\end{array} \right. \nn
Y_k = \bpm \cosh{\mu_{1k} \over 2} & 0 & 0 & \sinh{\mu_{1k}\over2} \\
0 & \cosh{\mu_{2k}\over2} & \sinh{\mu_{2k}\over2} & 0 \\
0 & \sinh{\mu_{2k} \over 2} & \cosh{\mu_{2k}\over2} & 0 \\
\sinh{\mu_{1k}\over 2} & 0 & 0 & \cosh{\mu_{1k} \over 2}
\epm,&\quad& \left\{ \begin{array}{cc} \sinh\mu_{1k} =
\frac{\Gamma_k}{E_{1k}}, & \quad\sinh\mu_{2k} =
\frac{\Gamma_k}{E_{2k}}
\\ {} \\ \cosh\mu_{1k} =
\frac{\mathcal{E}_{1k}}{E_{1k}}, & \quad\cosh\mu_{2k} =
\frac{\mathcal{E}_{2k}}{E_{2k}}
\end{array} \right. \ea
\ew where $\Lambda_{k} = \lambda_{k}\cos\theta_{k}$, $\Gamma_{k} =
\lambda_{k}\sin\theta_{k}$, $\Delta_{k} = \frac{\Delta_{1k} +
\Delta_{2k}}{2}$, $\delta_{k} = \frac{\Delta_{1k} -
\Delta_{2k}}{2}$, $\mathcal{E}_{1k,2k} = \sqrt{\Delta_{k}^{2} -
\Lambda_{k}^{2}} \pm \delta_{k}$, $E_{1k,2k} =
\sqrt{\mathcal{E}_{1k,2k}^{2} - \Gamma_{k}^{2}}$, and
$\Delta_{1k,2k} = \frac{\varepsilon_{k} + \omega}{2} \pm
\frac{1}{2}\sqrt{(\varepsilon_{k} - \omega)^{2} +
4\lambda_{k}^{2}}$.
The final form of the Hamiltonian is

\ba && Y^{\dag}_k X^{\dag}_k W^\dag_k  {\cal H}_{k} W_{k} X_k Y_k =
\mathrm{diag} ( E_{1k}, E_{2k}, E_{2k}, E_{1k} ) \nn
&& H = \sum_k \left( E_{1k} \calAbar_{1k} \calA_{1k} + E_{2k}
\calAbar_{2k} \calA_{2k}\right) .\ea
The sum $\sum_k$ is over the entire Brillouin zone. For any given
$k$ we have $E_{1k} \ge E_{2k}$, constituting an upper and lower
branch of the spectra.

\begin{figure}[t]
\begin{center}
\includegraphics[scale=0.5]{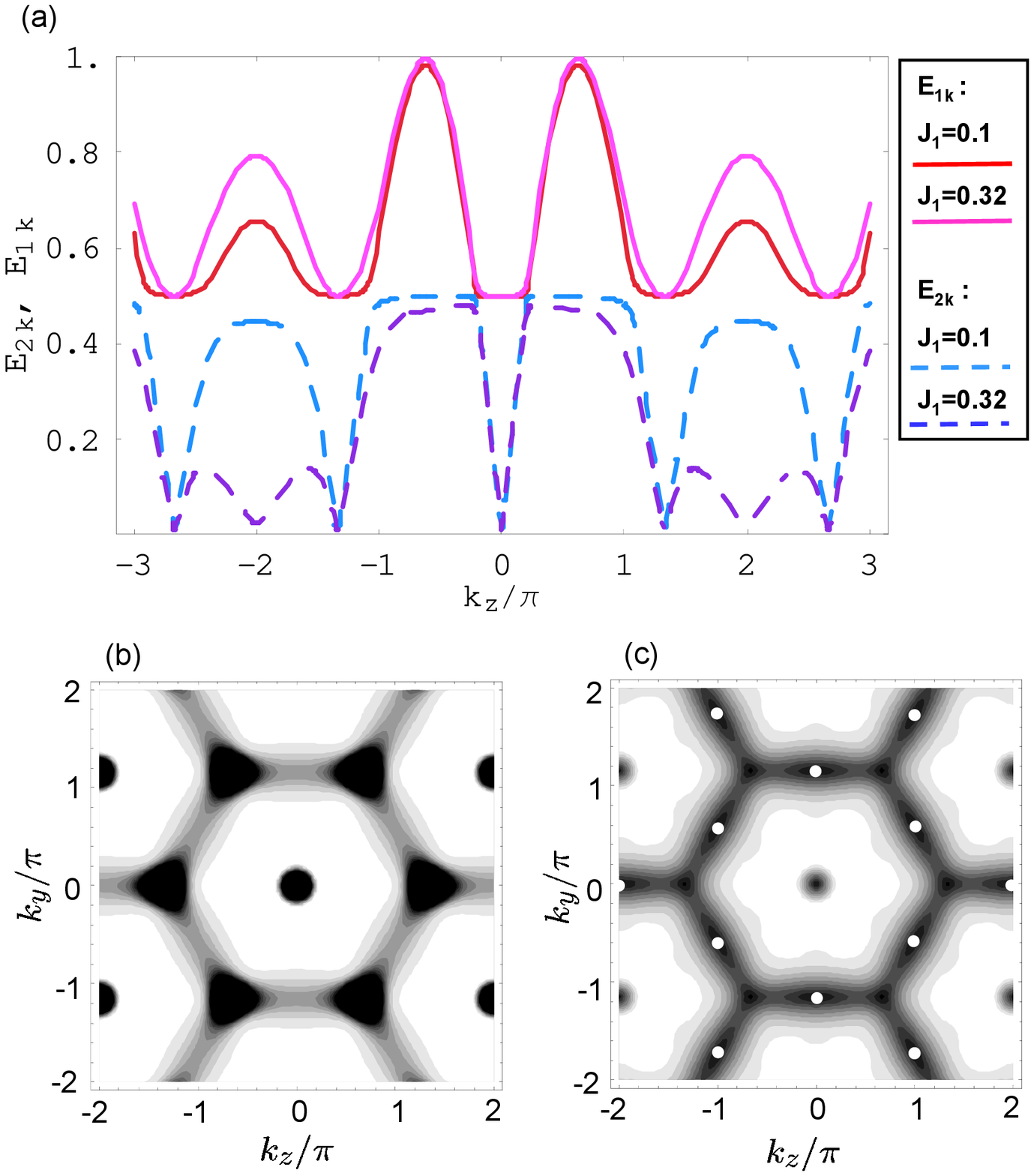}
\end{center}
\caption{(color online) (a) Dispersion along the $(k_{z},0)$
direction for $E_{1k}$(solid) and $E_{2k}$(dashed) for two choices
of spin-phonon coupling, $J_{1} = 0.1$ (red and blue lines) and
$J_{1} = 0.32$ (pink and violet lines). We have chosen $J_0 = 3.7$
to normalize the maximum energy value to one. Other parameters are
$S=1/2$, $\omega_0 =0.5$, and $m=2$. The phonon wave function width
is chosen $1/\sqrt{m\omega_0} = 1$, equal to the lattice constant.
The level repulsion is particularly severe at
$(k_{z},k_{y})=(2\pi,0)$ as the strength of the coupling is
increased. (b)-(c) Contour plots of the low-energy branch $E_{2k}$
for (b) $J_1 = 0.1$  and (c) 0.32. Indicated as white dots in (c)
are the $k$ points where $E_{2k}$ reaches zero at the critical
spin-phonon coupling. } \label{dispersion}
\end{figure}

A plot of $E_{1k}$ and $E_{2k}$ in Fig.~\ref{dispersion} shows the
change in the magnon and the phonon spectrum as $J_{1}$ is
increased. The most notable feature in the coupled energy spectrum
is the appearance of the roton-like minimum at a set of $k$-points
in the Brillouin zone. When $J_1$ equals the threshold value,
\textit{e.g.} $J_{1c}\equiv 0.321$ for $S=1/2$, $\omega_0 =0.5$, and
the phonon wave function width $m\omega_0 =1$, $E_{2k}$ touches zero
at $k = (2\pi,0)$, $(0,2\pi/\sqrt{3})$, and all their sixfold
symmetry-related points indicated as white dots in
Fig.~\ref{dispersion} (c). The original zeros of the magnon spectrum
at $\pm (4\pi/3,0)$ remain intact through nonzero $J_1$.

\begin{figure}[t]
\begin{center}
\includegraphics[scale=0.4]{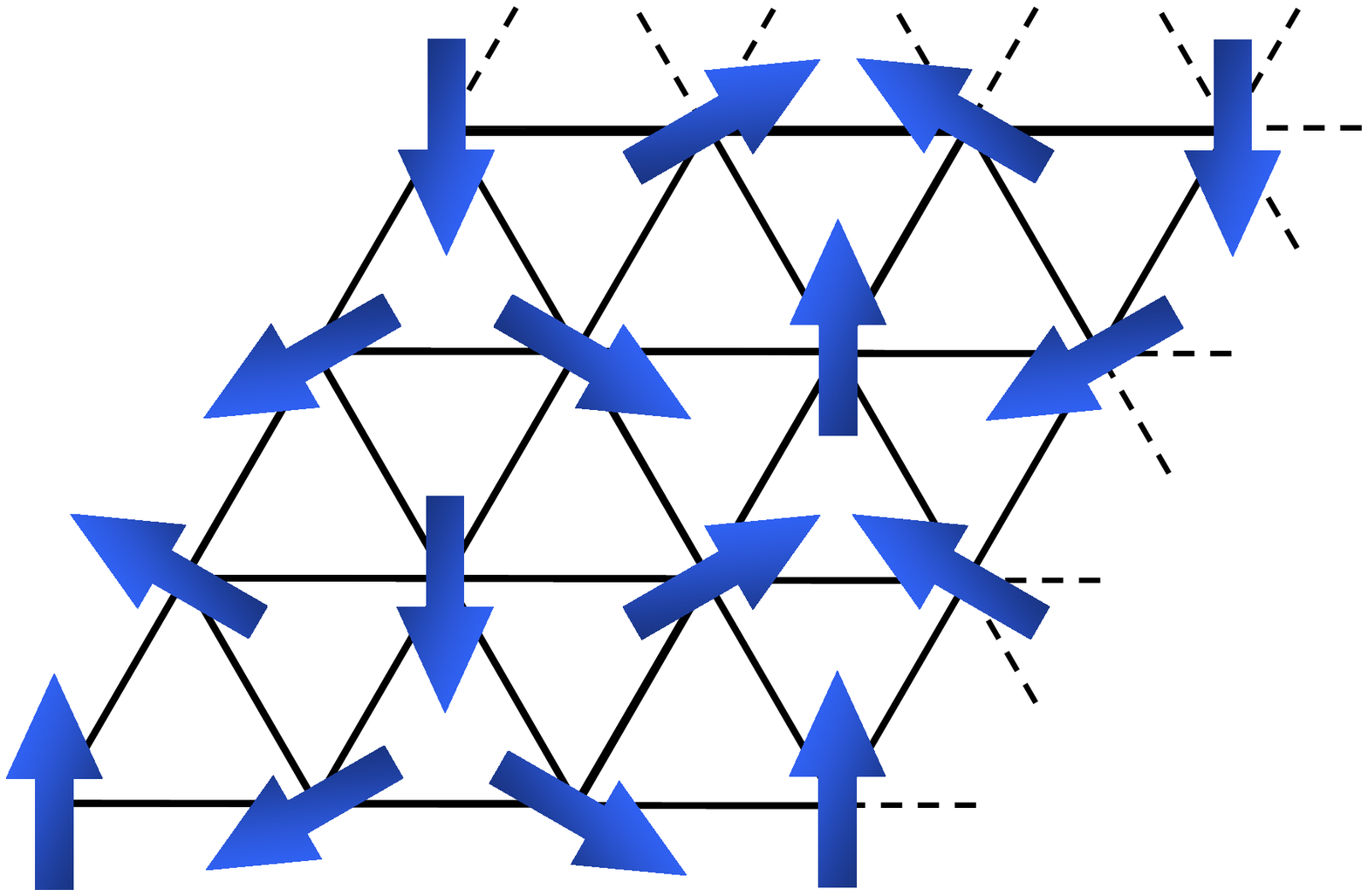}
\end{center}
\caption{(color online) The new emergent spin configuration for the
critical spin-phonon coupling value $J_1 =J_{1c}$ corresponding to
the ordering wave vector $Q'=(4\pi/3 \pm 2\pi,0)$, or, equivalently,
$Q'=\pm (2\pi/3,0)$. This configuration becomes degenerate with
those at $Q= \pm (4\pi/3,0)$ when $J_1$ equals $J_{1c}$.}
\label{newconfig}
\end{figure}

As this happens, one has a new spin-ordered pattern illustrated in
Fig.~\ref{newconfig} becoming degenerate with the original
120$^\circ$ ordered phase. This new pattern is obtained by rotating
spins counterclockwise by 120$^\circ$ along the $\hat{e}_{1}$
direction and by 60$^\circ$, also counterclockwise, along the
$\hat{e}_{2}$ direction. The state where the spins are rotated
clockwise in both directions will also be degenerate, carrying the
opposite sense of chirality. In terms of ordering wave vectors, the
new ground states are characterized by $Q' = \pm (2\pi/3,0)$ instead
of $Q = \pm (4\pi/3,0)$ as in the 120$^\circ$ ordered phase.

\section{Physical Quantities}\label{sec:physical quantities}

\begin{figure}[h]
\begin{center}
\includegraphics[scale=0.4]{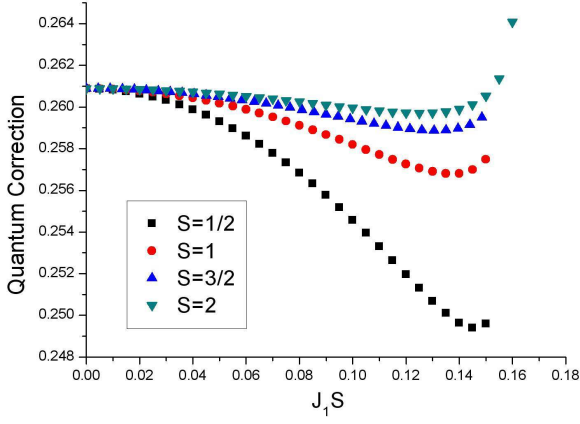}
\end{center}
\caption{(color online) Plot of the quantum correction versus the
coupling strength for various spin values and $0 < J_1 S<J_{1c} S$.
The critical coupling strength depends on $S$, while the product
$J_1 S$ is nearly independent of $S$. Choices of other parameters
are the same as in Fig.~\ref{dispersion}.} \label{qc}
\end{figure}

The local staggered magnetization (uniform magnetization in the
rotated basis), $ \langle \bm{S}\rangle  \equiv (1/N) \sum_{i} \la
\bm{S}_{i}^{z} \ra $, is modified due to the spin-phonon coupling.
The quantum correction, defined as the difference of the classical
and quantum averages $S - \langle \bm{S} \rangle $, reads
\ba \label{qmag} \sum_{k} \left( \sinh^{2}\frac{\phi_{k}}{2} +
\cosh\phi_{k} \la \alpha_{k}^{\dag}\alpha_{k} \ra -\sinh\phi_{k} \la
\alpha_{k}^{\dag}\alpha_{\mk}^{\dag} \ra \right), \ea
which is plotted at $T=0$ in Fig.~\ref{qc} for various spin values
as the coupling strength is varied. In small powers of $\lambda_k$
one obtains the following perturbative expression as the quantum
correction

\be \label{pert}  \sum_k \sinh^{2}\frac{\phi_{k}}{2} + \sum_k
{\lambda_{k}^2 \cosh \phi_k  \over (\varepsilon_{k} \!+\! \omega_0
)^{2} }\!+\!  \sum_k {\lambda_k^2 \sinh \phi_k \over \varepsilon_k
(\varepsilon_k + \omega_0 ) }. \ee
The first term is the usual quantum fluctuation correction, and the
latter two account for further corrections due to spin-phonon
coupling. There is only a tiny change in the local magnetization
affected by the spin-phonon coupling. On the other hand, the upturn
found in Fig.~\ref{qc} as $J_1$ is driven up to its critical value
is probably indicative of the diverging quantum correction as the
new ground state is approached. Due to the finite phonon mass
$\omega_0$, the spin-phonon coupling effects are not apparent until
very near the criticality.

\begin{figure}[h]
\begin{center}
\includegraphics[scale=0.45]{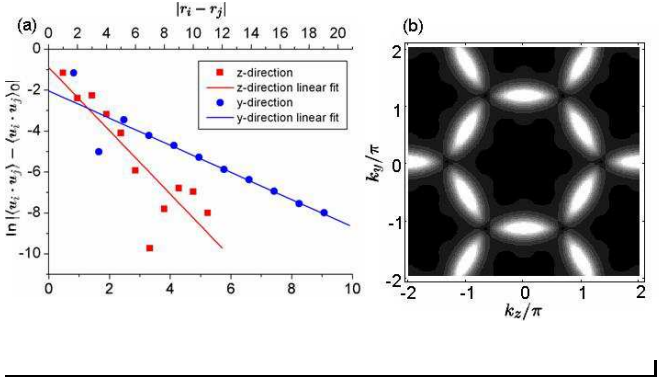}
\end{center}
\caption{(color online) (a) Log plots of the correlation function $|
\la \bm{u}_{i} \cdot \bm{u}_{j}\ra -\la \bm{u}_{i} \cdot
\bm{u}_{j}\ra_0 |$ in the $z$ and $y$ directions. Using the same
parameter values as in Fig.~\ref{dispersion}, the correlation length
can be extracted as $\sim 1.36$ and $\sim 3.16$ lattice constants in
the $z$ and $y$ directions, respectively. (b) Plot of $G_k -1$.
Bright regions indicate peaks in  $ G_k -1 $. } \label{fig3}
\end{figure}

The equal-time phonon correlation function $\la \v u_{i}  \cdot \v
u_{j} \ra$, which will be short-ranged $\la \v u_{i} \cdot \v u_{j}
\ra_0  = \delta_{ij}$ (since we have chosen $m\omega_0 = 1$) in the
absence of spin-phonon coupling, now reads

\ba && \la \v u_{i} \cdot \v u_{j} \ra \!-\!\la \v u_{i} \cdot \v
u_{j} \ra_0  = \frac{\sqrt{3}}{16\pi^{2}}
\int_{\mathclap{\mathrm{~~~BZ}}}
e^{ik\cdot(r_{i}\!-\!r_{j})}(G_{k}\!-\! 1)d^{2}k , \nn
&& ~~~~~G_k = \la ( \beta_{k} +\beta_{\mk}^{\dag} )(
\beta_{k}^{\dag} + \beta_{\mk} )\ra  , \label{correlation}\ea
at zero temperature. A log plot of  $| \la \v u_{i} \cdot \v u_{j}
\ra \!-\!\la \v u_{i} \cdot \v u_{j} \ra_0  |$ is given in
Fig.~\ref{fig3}, showing an exponential decay with a correlation
length which depends on the parameters. The momentum space
correlation $G_k$ shows pronounced peaks around $k=
(0,2\pi/\sqrt{3})$ and other symmetry-related points as shown in
Fig.~\ref{fig3} (b). These are the same points where $E_{2k}$ shows
pronounced minima for large $J_1$. Detection of such peaks in the
phonon structure factor $G_k$ will be instrumental in identifying
the spin-phonon coupling in a triangular antiferromagnet.

\begin{figure}[t]
\begin{center}
\includegraphics[scale=0.45]{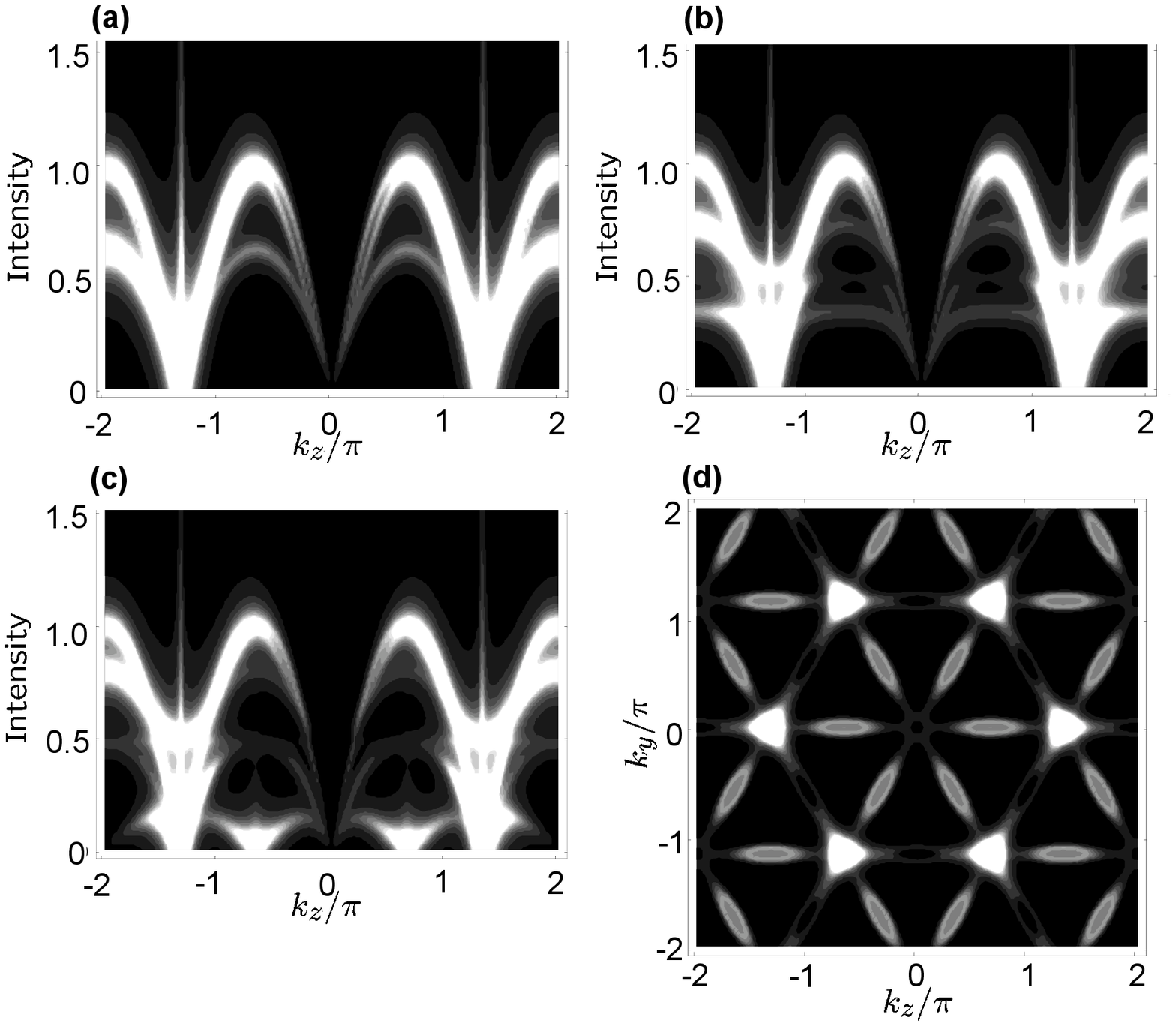}
\end{center}
\caption{Plots of the spectral function $I(k, \omega)$ along the
$k_{z}$-direction ($k_y = 0$) for (a) $J_{1} = 0$, (b) $J_{1} =
0.2$, and (c) $J_{1} = 0.32$. Emergence of a new low-energy feature
at $k_z = 2\pi/3$ for $J_1$ close to the critical value $J_{1c} =
0.321$ is apparent in (c). (d) Plot of $I(k_z, k_y, \omega =0.1)$
clearly indicates new intensity peaks at $(2\pi/3,0)$ and other
symmetry points (elongated and shaded) while the bright patterns at
$(4\pi/3,0)$ and etc. are due to the original spin waves. }
\label{spectralfunction}
\end{figure}

The spin-spin correlation function, $\la S_{j}^{+}(t) S_{i}^{-}(0)
\ra$, can be an effective probe of the spin-phonon coupling. Using
the straightforward algebra we have calculated the absorption
spectra $I (k, \omega)$ as the imaginary part of the Fourier
transform of the spin-spin correlation function,
\begin{widetext}
\ba \label{ikw} I(k,\omega) &=& \frac{\pi S}{8} \bigg[
\sum_{\alpha=1,2} e^{\phi_{k_{\alpha}}}\Big( ( B_{1k_{\alpha}} -
A_{1k_{\alpha}} )\delta( \omega - E_{1k_{\alpha}} ) +
(B_{2k_{\alpha}} - A_{2k_{\alpha}})\delta( \omega - E_{2k_{\alpha}} )
\Big)\nn && \ph{~~\frac{S\pi}{8}} + 2e^{-\phi_{k}}\Big( (A_{1k} +
B_{1k})\delta( \omega - E_{1k} ) + (A_{2k} + B_{2k})\delta( \omega -
E_{2k} ) \Big) \bigg],
\ea
\end{widetext}
which is plotted in Fig.~\ref{spectralfunction}. The functions
appearing in  Eq.~\eqref{ikw} are defined as $k_{1,2}= k\pm
(4\pi/3,0)$, and
\ba \label{spinspin} A_{1k} &=& \sinh\mu_{1k}(\cos\theta_{k} +
\cosh\nu_{k}) - \cosh\mu_{1k}\sin\theta_{k}\sinh\nu_{k}\nn
A_{2k} &=& \sinh\mu_{2k}(\cos\theta_{k} - \cosh\nu_{k}) -
\cosh\mu_{2k}\sin\theta_{k}\sinh\nu_{k}\nn
B_{1k} &=& \cosh\mu_{1k}(\cos\theta_{k} + \cosh\nu_{k}) -
\sinh\mu_{1k}\sin\theta_{k}\sinh\nu_{k}\nn
B_{2k} &=& \cosh\mu_{2k}(\cosh\nu_{k} - \cos\theta_{k}) +
\sinh\mu_{2k}\sin\theta_{k}\sinh\nu_{k}. \nn \ea
The flattening and the eventual collapse of the excitation band
found earlier now manifests itself as intensity patterns at $(k_z,
k_y ) = (2\pi/3,0)$ and its rotational counterparts as can be seen
in Fig.~\ref{spectralfunction} (d).

\section{Discussion}\label{sec:discussion}

In summary, we have considered the magnon-phonon coupling in the
exchange-striction coupled triangular lattice antiferromagnet for
Heisenberg spins in the Holstein-Primakoff approach. The dynamics of
the lattice and the spins are coupled to produce interesting
modifications in the excitation spectra, in particular (i) the
significant lowering of the magnon excitation energy at wave vector
transfer $\pm 2\pi$ as indicated in Fig.~\ref{dispersion}, and (ii)
the concordant variation in the phonon structure factor as shown in
Fig.~\ref{fig3} (b).  Detection of an additional low-energy spectra
in the spin spectral function $I(k,\omega)$ and in the phonon
structure factor $G_k$ through neutron scattering experiments will be
a clear hint of the strong spin-phonon coupling.

Naively speaking, integrating out the phonon coordinate from
Eq.~\eqref{our-model} would generate the effective interaction, $\sim
-\sum_{i}(\sum_{j =\textrm{NN of $i$}} \hat{e}_{ij} S_{i}\cdot
S_{j})^{2}$, which embodies the ferromagnetic biquadratic exchange,
$-(S_{i}\cdot S_{j})^{2}$, and some three-body interactions as well.
A quantum spin model involving quadratic and biquadratic exchanges
were considered extensively\cite{spin-nematic} following the
discovery of the liquid-like ground state in the triangular
antiferromagnet NiGa$_2$S$_4$\cite{nakatsuji1}. The ground state
revealed correlations, dynamic on the scale of $\sim$~1 ns, of a
period-six spin orientation ($60^{\circ}$ angles between nearest
neighbors), quite unlike the period-three orientation ($120^{\circ}$
angles between nearest neighbors) expected in the triangular lattice.
It is not easy to reproduce such a spin structure in the mean field
solution of the spin models considered in
Refs.~\onlinecite{spin-nematic}. In fact, the spin-spin correlation
observed in NiGa$_2$S$_4$ is almost exactly the one shown in
Fig.~\ref{newconfig}. To correctly account for the observed
periodicity of spins in NiGa$_2$S$_4$, one would have to consider an
extended-neighbor interaction as in Ref.~\onlinecite{nakatsuji2}, or
some dynamical consequence of spin-phonon coupling as in the present
paper.

We wish to acknowledge fruitful discussions with Chenglong Jia.
Discussions with Satoru Nakatsuji on NiGa$_2$S$_4$ are also
gratefully acknowledged. H. J. H. was partly supported by the Samsung
Research Fund, Sungkyunkwan University, 2006.


\begin{thebibliography}{99}
\bibitem{comment} In the unrotated spin basis the phonon at momentum
$k$ couples to the magnon at $k\pm Q$, where $Q$ is the ordering
wave vector $(4\pi/3,0)$. Due to the rotation that aligns the spins
ferromagnetically, the magnon-phonon coupling appears between the
same $k$ states.
\end{thebibliography}
\end{document}